\documentclass[twocolumn,twoside]{article}

\usepackage[T2A,T1]{fontenc}
\usepackage[english]{babel}
\usepackage{abstract}
\usepackage{a4}
\usepackage{bm}
\usepackage{amssymb}
\usepackage[numbers,square,sort&compress]{natbib}
\usepackage{hyperref}

\makeatletter
\renewcommand{\@seccntformat}[1]{\csname the#1\endcsname.\,\,\,\,
  \expandafter\ifx\csname c@#1\endcsname \c@subsubsection
    \!
  \fi
}
\makeatother
\usepackage[hang,flushmargin]{footmisc}
\usepackage{tocloft}

\usepackage{caption}
\begin{document}

\title{\textbf{On variational trial functions in the extended Thomas-Fermi method}}

\author{A.~Y. Potekhin$^1$, 
 A.~I. Chugunov$^1$, N.~N. Shchechilin$^2$, N. Chamel$^2$}

\renewcommand{\abstractname}{}
\date{\parbox{.95\linewidth}{\small$^1$Ioffe Institute, Politekhnicheskaya 26,
194021 Saint Petersburg, Russian Federation
\\
$^2$Institut d'Astronomie et d'Astrophysique, CP-226,
Universit\'e Libre de Bruxelles, 1050 Brussels, Belgium}}

\twocolumn[
    \begin{onecolabstract}    
\maketitle
\normalsize\noindent{Parametrized nucleon density distributions are widely employed for the 
calculation of the properties of atomic nuclei and dense inhomogeneous 
matter in compact stars within the Thomas-Fermi method and its 
extensions.  
We show that the use of insufficiently
smooth parametrizations may deteriorate the accuracy of this method. We
discuss and clarify the smoothness condition  using the example of the
so-called ``nuclear pasta'' in the neutron star mantle.}\\
\phantom{blabla}
\\
\phantom{blabla}
\end{onecolabstract}
]
\saythanks 

\thispagestyle{empty}




\section{Introduction}

Thomas \cite{Thomas27} and Fermi \cite{Fermi27} suggested a statistical
description of an atom with a large number of electrons. They used
the local electron number density to calculate the potential energy and chemical
potential of an electron in the self-consistent field of other electrons
and the nucleus. Subsequently the Thomas-Fermi (TF) method was refined
by various corrections and applied for studying a large variety of
many-body systems and dense matter (see, e.g.,
Refs.\,\cite{KirzhnitsLS75,RingSchuck80,BrackBhaduri97,Shpatakovskaya12}, for
reviews). We will focus on the extended Thomas-Fermi (ETF)
theory, characterized by inclusion of the so called gradient
corrections,
which are functions of density derivatives.

The first extension to the TF model was proposed by von Weizs\"acker
\cite{Weizsacker35} for the description of heavy atomic nuclei. He
introduced a gradient correction  to the kinetic energy of a system of
nucleons, in order to capture the surface effect on the nuclear binding
energy. Kompaneets and Pavlovskii \cite{KompaneetsPavlovskii57} showed
that the leading-order quantum correction to the TF model actually
equals 1/9 of the von Weizs\"acker gradient correction. Kirzhnits
\cite{Kirzhnits57} introduced a regular perturbative method for the
derivation of corrections in powers of gradient operator $\nabla$
applied to an effective potential. Note that this approach leads to
appearance of quantum corrections for the energy $E$ as well as for the
number density ${n}$. Hodges \cite{Hodges73} represented the Kirzhnits
method in a more straightforward way, derived, for the first time, the
correct explicit form of the corrections to power 4 in  $\nabla$, and
showed that the quantum corrections to the density can be excluded from
explicit consideration by writing the final expression for the energy in
terms of the operator $\nabla$ applied to the particle number density
(see also \cite{Grammaticos_Voros79,BrackGH85}). 

The gradient corrections can be derived from the Wigner-Kirkwood
expansion of the Bloch density matrix around its value obtained in the
TF approximation in powers of the reduced Planck constant $\hbar$ (see,
e.g., Refs.\,\cite{RingSchuck80,BrackBhaduri97}). The lowest order terms
of the ETF model reproduce the TF expressions, and the remaining
expansion includes only even powers of $\hbar$. The lowest-order
corrections ($\propto\hbar^2$) involve the terms with $(\nabla {n})^2$
and $\nabla^2 {n}$. Going up to order $\hbar^4$ in the expansion, one
obtains the next correction, containing up to the fourth derivatives of
${n}$. 

Beside the gradient corrections, other refinements of the TF model
include shell corrections, exchange and correlation
effects
\cite{KirzhnitsLS75,BrackBhaduri97,Shpatakovskaya12}, and a pairing
interaction correction for the description of many-nucleon systems
\cite{BengtssonSchuck80}. In this context, Brack et al.~\cite{Brack84,BrackGH85}
developed a theory, named TEFT, which generalizes the ETF model to 
finite temperatures~$T$.

It is worth saying more about the description of many-nucleon systems. Unlike
plasma, consisting of electrons and atomic nuclei, which under laboratory
conditions can be considered as point particles interacting according
to Coulomb law, in many-nucleon systems the main role is played by strong
interactions between nucleons
having finite sizes and consisting of quarks. In
a plasma, as a rule, the Born-Oppenheimer approximation is applicable,
in which electron dynamics is calculated neglecting 
motion of nuclei, whereas in the many-nucleon systems,
protons and neutrons act
as two equal kinds of particles.
Moreover, in order to correctly
describe the properties of relatively light nuclei, it is necessary
to add a three-particle potential to the two-particle interaction potential
determined from nucleon scattering experiments 
\cite{Pudliner_95,ArriagaPW95,Wang_ea24_3n_required}.
For
the theoretical description of the properties of heavier atomic nuclei and
dense matter, as a rule,
self-consistent field methods with phenomenological \textit{effective}
potentials of two-particle nucleon-nucleon interactions are used,
designed in such a way as to simulate
many-particle effects as well (see Ref.~\cite{Bender_HR03} for review).
The effective
potentials that make it possible to reproduce the properties of atomic
nuclei have a complex form. A well-known class of such potentials is
the Skyrme potential \cite{Skyrme59} and its modifications (for example,
\cite{ChamelGP09}). Hartree-Fock equations with Skyrme interactions
for spherical nuclei were first derived in Ref.~\cite{VautherinBrink72}. They have the form of
Schr\"odinger equations
for single-particle wave functions, which include
\emph{effective masses} of nucleons $M^*$ and self-consistent potentials
depending on the particle number density, kinetic energy density,
and spin-orbital density.

The nucleons in a dense matter can form Cooper pairs and become
superfluid. The most accurate self-consistent field method for such
matter is the Hartree-Fock-Bogoliubov (HFB) method (see reviews
\cite{RingSchuck80,Bender_HR03} and references therein). It can be used
to describe heavy atomic nuclei (including neutron-rich nuclei in the
outer crust of neutron stars). However, if nuclei, or rather nucleon
clusters, are embedded in highly degenerate neutron matter (for
instance, in the inner crust of neutron stars), the HFB method becomes
too expensive. In this case, the HFB method can be fairly accurately
approximated using the computationally much faster ETF method with
consistent shell and  pairing corrections added perturbatively at the
end of the ETF calculation
\cite{Onsi_08,Pearson_12,Pearson_15,ChamelPS24}
(see \cite{ShelleyPastore20} for detailed comparisons with the HFB
method).

A semi-phenomenological TF theory of atomic nuclei, which included a
second-order gradient correction with an empirically adjusted
coefficient, was developed by Bethe~\cite{Bethe68} (his paper gives also
basic references to preceding models). Brack et al.~\cite{BrackJC76}
found that the fourth-order ETF theory does not need such empirical
adjustment to reproduce the nuclear binding energies to within a few
MeV, while the inaccuracy reached tens of MeV if only the second-order
corrections were included. The modern ETF theory, which  relies on
accurately calibrated effective nucleon-nucleon interactions and
includes Strutinsky shell corrections \cite{Strutinsky67,Jennings73} and
pairing corrections \cite{BengtssonSchuck80,RingSchuck80}, allows one to
reproduce all measured masses of heavy nuclei with typical errors
$\sim0.7$~MeV (e.g., \cite{Aboussir_92}; see Ref.~\cite{Bender_HR03} for
review and references).

In the following, we will consider applications of the ETF model to the
exotic states of dense matter, where the nucleon clusters are not
necessarily quasi-spherical, but may also take the shapes of cylinders
(dubbed ``spaghetti'') and plates (``lasagna'')
\cite{RavenhallPW83,LorenzRP93}.  In addition to these ``pasta phases''
with nucleon clusters  immersed in a more dilute background of free
nucleons (mainly neutrons)  and electrons,  there may exist ``inverse''
pasta phases with localized nucleon  depletions in the dense matter
\cite{PethickRavenhall95}, called ``anti-spaghetti''  or ``bucatini'' 
(inverted cylinders) and ``Swiss cheese'' (inverted spheres)
\cite{HaenselPY07,ChamelHaensel08,Schneider_13}. The pasta phases may be
in thermodynamic equilibrium (the ground state) at mean  nucleon number
densities  $\bar{n}\sim 0.05$ fm$^{-3} -0.08$ fm$^{-3}$. The nuclear
pasta layers are called the mantle; they are located between  the solid
crust and the liquid core of a neutron star (see, e.g.,
\cite{HaenselPY07,ChamelHaensel08}, and references therein). In terms of
its elastic properties, the mantle is close to liquid crystals, as shown
in Ref.~\cite{PethickPotekhin98} (see
\cite{Pethick_ea20,Xia_ea23,ZemlyakovChugunov23} for the current state of the theory of
mantle elasticity).

The densities in the mantle are only 2--3 times lower than the nuclear
saturation density  $n_\mathrm{sat}$, which corresponds to the zero
pressure in the model of uniform symmetric nuclear matter and is close
to the typical number density of the nucleons in heavy atomic
nuclei.\footnote{Estimates of the saturation density vary from 0.14
fm$^{-3}$ \cite{RingSchuck80} to $0.17$ fm$^{-3}$
\cite{HaenselKP81,Sakuragi16}. The value $n_\mathrm{sat} = 0.16$
fm$^{-3}$ is most often quoted, based on the study by Tondeur et
al.~\cite{TondeurBF86}. Recent experimental results suggest
$n_\mathrm{sat}=0.15\pm0.01$ fm$^{-3}$ \cite{HorowitzPR20}.}

The TF model  and its refined versions are  often combined with the
Wigner-Seitz (WS) approximation \cite{WignerSeitz33,Salpeter61}, which
reduces the problem to consideration of a single WS cell, replaced by an
overall neutral sphere of the same volume (see Ref.~\cite{Chamel_07} for
a discussion of limitations of this approximation).  Analogously for the
phase with cylindrical nuclei, one considers an overall neutral
cylinder. In the case of plate-like nuclei, the true WS cell is a slab,
so that a further geometrical simplification is unnecessary.

The ground state of dense
matter corresponds to the minimum of energy $E=
\int_V \mathcal{E}\,{d}V$ for a given nucleon number ${N}\equiv \int_V
(n_{p}+n_{n})\,{d}V$ in volume $V$ under the charge neutrality
constraint  $\int_V (n_{p}-n_{e})\,{d}V = 0$, where $n_{e}$, $n_{p}$,
and $n_{n}$ are the local number densities of electrons, protons, and
neutrons, respectively, and $\mathcal{E}$ is the energy density, which 
can be conventionally decomposed as
\begin{equation}
   \mathcal{E} = \mathcal{E}_\mathrm{nuc} + \mathcal{E}_{e} +
   \mathcal{E}_\mathrm{Coul} + \mathcal{E}_\mathrm{cor}.
\end{equation}
Here, $\mathcal{E}_\mathrm{nuc}$ and $\mathcal{E}_{e}$ are contributions
from the nucleons and electrons, respectively,
$\mathcal{E}_\mathrm{Coul}$ denotes energy density due to Coulomb
interactions (beyond those already accounted in
$\mathcal{E}_\mathrm{nuc}$ and $\mathcal{E}_{e}$), and
$\mathcal{E}_\mathrm{cor}$ stands for various corrections, which may
appear in the refined theory, such as shell and pairing corrections. The
nuclear energy density $\mathcal{E}_\mathrm{nuc}$ can be written in the
nonrelativistic approximation as (e.g., Ref.\,\cite{OnsiPP97})
\begin{equation}
   \mathcal{E}_\mathrm{nuc} = \sum_q \frac{\hbar^2}{2M_q^*}\,\tau_q
     + \mathcal{V},
\label{Edens}
\end{equation}
where $M_q^*$ is a density-dependent effective mass of a nucleon of type
$q$   ($q={n}$ for neutrons and $q={p}$ for protons), $\tau_q$ is a
normalized kinetic energy of a nucleon, generated by the momentum square
operator in the Hamiltonian, and $\mathcal{V}$ stands for a sum of
potential-energy terms. In the ETF model, both terms in (\ref{Edens})
depend on local number densities ${n}_q(\bm{r})$, their gradients
$\nabla{n}_q(\bm{r})$, and spin current densities $\bm{J}_q(\bm{r})$
with coefficients determined by an employed realization of the
Skyrme-type potential. The gradient expansion enables one to express 
$\tau_q(\bm{r})$ and $\bm{J}_q(\bm{r})$ (and therefore
$\mathcal{E}_\mathrm{nuc}$) as functions entirely of the nucleon
densities ${n}_q(\bm{r})$ and their derivatives \cite{OnsiPP97}. 

The energy minimum can be found  by solving the corresponding
Euler-Lagrange (EL) equations.  For example, Barkat et
al.~\cite{BarkatBI72} numerically solved the EL equations derived from the
semi-phenomenological theory of Bethe~\cite{Bethe68}. Later the EL
equations were solved by different authors in the TF and ETF models for
finite nuclei \cite{SuraudVautherin84,Suraud87,Centelles_90} and nuclear
matter \cite{SuraudVautherin84,Suraud85,BartelBD85,CentellesDEV98} using
the so called imaginary time-step method \cite{Davies_80}.  However, the
fourth-order ETF theory, when used in conjunction with Skyrme forces 
(see, e.g.,  Ref.~\cite{Bender_HR03} and references therein).
leads to complex, highly non-linear EL equations for the particle
densities, which are difficult to solve exactly
\cite{BartelBD85,Centelles_90}. To avoid these difficulties and speed up
computations, the ETF energy functional is most often minimized
explicitly within a family of \textit{parametrized} 
nucleon
density profiles
$n_{n}(\bm{r})$ and $n_{p}(\bm{r})$ (examples will be given below). This
\emph{restricted variational method} drastically simplifies and accelerates
calculations. At the same time, it can be very accurate, as was
demonstrated by Bartel et al.~\cite{BartelBD85}.

The ETF method is subject to different kinds of divergences. Being based
on expansions in powers of $\hbar$ around the classical values,  the
gradient corrections can only be calculated inside the classically
allowed region of particle motion. In the case of finite or semi-finite
classical orbits, the semiclassical expansions of the  number density
${n}(\bm{r})$ and kinetic energy density $\tau(\bm{r})$ are not defined
at the classical turning points and beyond, and they diverge near these
points. However, this is not a problem, because these divergences are
integrable and allow one to obtain non-diverging expressions of the
gradient corrections to $\tau$ as functions of ${n}$
\cite{Grammaticos_Voros79}. Thus the ETF number densities can be understood as
generalized distributions with well-defined integrals and moments
\cite{RingSchuck80,BrackGH85}. It is also of interest that a non-zero
temperature in the TEFT model removes the divergence of the ETF
densities and 
allows their extension to the classically forbidden regions \cite{Brack84}. 

Other divergences are related to the asymptotic nature of the gradient
expansion. The sixth- and higher-order gradient terms diverge for
densities which fall exponentially to zero as functions of $\bm{r}$, and
must therefore be left out \cite{Brack84,BrackGH85}, at least in
calculations of the properties of isolated atoms or nuclei. Thus only
the terms up to the fourth order constitute the converging part of the
ETF expansion.

However, this part can still diverge 
within 
the restricted
variational method, if the trial density distributions ${n}_q(\bm{r})$
are not sufficiently smooth. The goal of the present paper is to clarify
the smoothness conditions for the applicability of this method to the
fourth-order ETF theory in 
spherical, cylindrical, and plane-parallel WS cells.

In Sec.~\ref{sect:ETF} we recall the basic ETF expressions for the
kinetic energy of nucleons. In Sec.~\ref{sect:param}, we review the most
common parametrizations of the local density of nucleons, which were
employed in the variational calculations of the atomic nuclei and
non-uniform 
matter in the crust and mantle of neutron stars.
In Sec.~\ref{sect:divergence} we discuss the divergences of the ETF
energy functionals for different kinds of kinks of the parametrized
density distribution. In Sec.~\ref{sect:acc_lim} we derive an associated
accuracy limit for numerical discrete-mesh ETF calculations.
In Sec.~\ref{sect:concl}, conclusions are summarized.


\section{ETF energy density}
\label{sect:ETF}

The nuclear energy in volume $V$ is given by the integral
\begin{equation}
   E_\mathrm{nuc} = \int_V \mathcal{E}_\mathrm{nuc}\,{d} V,
\label{Enuc}
\end{equation}
where $\mathcal{E}_\mathrm{nuc}$ is given by Eq.~(\ref{Edens}).
The kinetic energy contribution in Eq.~(\ref{Enuc}) can be written as 
\begin{equation}
\mathcal{T}_q = \int_V \frac{\hbar^2}{2M_q^*}\tau_q\,{d} V ,
\quad
\tau_q = \tau_q^{(0)} + \tau_q^{(2)} + \tau_q^{(4)},
\label{Tq}
\end{equation}
 where 
$\tau_q^{(0)}$ is the TF contribution, while $\tau_q^{(2)}$ and
$\tau_q^{(4)}$ are the second and fourth order gradient corrections, 
respectively. 

For simplicity, let us  neglect the difference of the effective mass
$M_q^*$ from the bare nucleon mass $M_q$ and omit the spin-orbit terms. 
Then in the zero temperature limit \cite{BrackJC76,Grammaticos_Voros79}
\begin{eqnarray}
    \tau_q^{(0)} &=& 
     \frac35\,\left(3\pi^2\right)^{2/3}\,{n}_q^{5/3}, 
\label{L0}
\\
    \tau_q^{(2)} &=&  \frac13\,\nabla^2{n}_q
          + \frac{1}{36}\,\frac{(\nabla{n}_q)^2}{{n}_q}, 
\label{L2}
\\
    \tau_q^{(4)} &=&  \frac{{n}_q^{1/3}}{(3\pi^2)^{2/3}} \left\{
     \frac{1}{180}\,\frac{\nabla^4{n}_q}{{n}_q}
     -\frac{1}{72}\,\frac{\nabla{n}_q\cdot\nabla(\nabla^2{n}_q)
           }{{n}_q^2} 
     \right. 
\nonumber\\ &&
     \left.
     -\frac{7}{1080}\,\frac{(\nabla^2{n}_q)^2
           }{{n}_q^2} 
     -\frac{7}{2160}\,\frac{\nabla^2(\nabla{n}_q)^2
           }{{n}_q^2} 
       \right.
\nonumber\\ &&
     \left.
     +\frac{7}{324} 
       \frac{(\nabla{n}_q)^2 \nabla^2{n}_q
           }{{n}_q^3}
     +\frac{23}{810} 
       \frac{(\nabla{n}_q \cdot \nabla)^2\,{n}_q
           }{{n}_q^3}
     \right. 
\nonumber\\ &&
     \left.
      - \frac{1}{45} 
       \frac{(\nabla{n}_q)^4
           }{{n}_q^4}
       \right\}.
\label{T4}
\end{eqnarray}
The fourth and third derivatives of ${n}_q(\bm{r})$ can be eliminated
from
$E_\mathrm{nuc}$ by partial integration with the use of the
Ostrogradsky-Gauss theorem \cite{Hodges73,BrackGH85}.
For example, the second term in Eq.\,(\ref{T4}) can be transformed as follows:
\begin{eqnarray}&&
  \int_V \phi\nabla{n}_q\cdot\nabla(\nabla^2{n}_q)\,{d} V
  =
   \int_V \Big\{ \nabla\cdot \left[ \left( 
   \phi\,\nabla^2{n}_q \right)\nabla{n}_q \right]
\nonumber\\ &&\qquad 
   - \frac{{d}\phi}{{d}{n}_q}\,(\nabla{n}_q)^2\,\nabla^2{n}_q
    - \phi\,(\nabla^2{n}_q)^2 \Big\}\,{d} V
\nonumber\\&&
 \quad = \oint_{S} \left( 
   \phi\,\nabla^2{n}_q \right)\nabla{n}_q \cdot\mathbf{n}\,{d}{S}
\nonumber\\ &&\qquad
  -
    \int_V \left[ \frac{{d}\phi}{{d}{n}_q}\,(\nabla{n}_q)^2\,\nabla^2{n}_q
    + \phi\,(\nabla^2{n}_q)^2 \right]\,{d} V,
\label{Iparts}
\end{eqnarray}
where $\mathbf{n}$ is the outer normal to the surface $S$ of the considered
domain $V$,
and we have denoted $\phi\equiv -(1/72)(3\pi^2)^{-2/3}{n}_q^{-5/3}$
for brevity.
This method gives
\begin{equation}
    \mathcal{T}_q^{(4)} \equiv \frac{\hbar^2}{2M_q} \int_V \tau_q^{(4)}(\bm{r})\,{d} V
      =\mathcal{T}_{q,v}^{(4)}+\mathcal{T}_{q,s}^{(4)}, 
\label{Tq4}
\end{equation}
where
$\mathcal{T}_{q,v}^{(4)}$ and $\mathcal{T}_{q,s}^{(4)}$ can be written as
\begin{equation}
    \mathcal{T}_{q,v}^{(4)} \equiv \frac{\hbar^2}{2M_q} \int_V
    \tau_{q,v}^{(4)}(\bm{r})\,{d} V,
\quad
    \mathcal{T}_{q,s}^{(4)} \equiv \frac{\hbar^2}{2M_q} \oint_S
    \tau_{q,s}^{(4)}(\bm{r})\,{d} S,
\end{equation}
and $\tau_{q,v}^{(4)}(\bm{r})$ includes only first- and second-order
derivatives of number density. Its explicit form is \cite{OnsiPP97}
\begin{eqnarray}
    \tau_{q,v}^{(4)} &=&  \frac{{n}_q^{1/3}}{(3\pi^2)^{2/3}} \left\{
     \frac{1}{270}\left(\frac{\nabla^2{n}_q}{{n}_q}\right)^2
     \right. 
\nonumber\\ &&
     \left.
     -\frac{1}{240}\,\frac{\nabla^2{n}_q}{{n}_q} 
       \left(\frac{\nabla{n}_q}{{n}_q}\right)^2
     +\frac{1}{810} 
       \left(\frac{\nabla{n}_q}{{n}_q}\right)^4
       \right\}\,,
\qquad
\label{L4}
\end{eqnarray}
which is equivalent to Eq.\,(30) of Hodges~\cite{Hodges73}. Usually one
assumes $\mathcal{T}_{q,s}^{(4)}=0$, so that $\tau_q^{(4)}$ can be
replaced by $\tau_{q,v}^{(4)}$ in Eq.~(\ref{Tq}). To this end,
Hodges~\cite{Hodges73} and Brack et al.~\cite{BrackGH85} integrated over
the whole space
considering in the context of an isolated atomic nucleus in vacuum 
 that ${n}_q(\bf{r})$ vanishes at $r\to\infty$
together with its derivatives. 
In the WS approximation, one can drop the
surface term $\mathcal{T}_{q,s}^{(4)}$, for example, under the
condition that the normal density
derivative to the WS cell surface $S$ equals zero\footnote{For instance,
Onsi et al.~\cite{OnsiPP97} stressed: ``it should be noted that the
fourth-order expressions are valid only on integrating over the whole of
space, or more generally, over a region on the surface of which the
density gradients vanish.''}:
\begin{equation}
 \mathbf{n}\cdot\nabla{n}_q(\bm{r})\big|_{\bm{r}\in S}=0.
\label{gradS}
\end{equation}
An analogous boundary condition was used by Wigner and
Seitz~\cite{WignerSeitz33} for electron wave functions. It naturally
follows from the symmetry and periodicity of ${n}_q(\bm{r})$
under the assumption that $\nabla{n}_q$ is continuous.

The continuity of $\nabla{n}_q$ was previously shown for the density
distributions that provide the exact minimum of $E_\mathrm{nuc}$ within
the fully variational Euler-Lagrange approach~\cite{BarkatBI72}. We will
see that the same continuity may be also required for parametrized
nucleon density distributions within the restricted variational
approach, in order to ensure convergence of the volume integral in
Eq.~(\ref{Enuc}).


\section{Parametrizations of nucleon density distributions}
\label{sect:param}

When matter is strongly compressed so that the mean nucleon number
density $\bar{n}\equiv {N}/V$ exceeds the neutron-drip density
$n_\mathrm{drip}\sim(2-3)\times10^{-4}$ fm$^{-3}$, neutrons start to
drip out of nuclei. Such huge densities occur in the inner crust of
neutron stars, where the nuclei  are immersed in a ``sea'' of unbound
neutrons.  One of the most  microscopic and tractable  methods of
describing such a dense matter so far  is the ETF theory with the shell
corrections for the protons (which remain bound in the nuclei) and with
the pairing corrections.

We will consider spherical, cylindrical, and slab-like WS cells and
assume that ${n}_q$ is symmetric with respect to the rotations around
the center of the sphere or  the axis of the cylinder, or to the
reflection with respect to the central plane of the slab. It means that
the density ${n}_q$ depends only on radial coordinate $r$  or  $r_\perp$
in the first and second cases,   respectively,  and only on $|z|$ in the
third case, where $z$ is the coordinate measured in the normal direction
from the central plane of the cell. The phases with spherical,
cylindrical, and plane-parallel symmetry are often called \mbox{three-},
\mbox{two-}, and one-dimensional  (3D, 2D, and 1D) structures,
respectively. However, one should keep in mind that such notation
concerns only the type of the symmetry of the density distribution,
whereas motion of particles (electrons and nucleons) remains
three-dimensional. These ``2D'' and ``1D'' structures should not be
confused with the true 2D and 1D systems, which were considered, e.g.,
in Ref.~\cite{Shpatakovskaya12} and which are described by the ETF
equations that differ from Eqs.~(\ref{L0})--(\ref{T4}) and (\ref{L4}). 

It is convenient to write a symmetric parametrization of the density
distribution in a WS cell in the form
\begin{eqnarray}
  {n}_q(\xi) &=& {n}_q^\mathrm{out} + {n}_{\Lambda q} \hat{f}_q(\xi),
\label{nLambda}
\\
  \hat{f}_q(\xi) & \equiv & \frac{ f_q(\xi) - f_q(R) }{ f_q(0) - f_q(R)},
\label{fnorm}
\end{eqnarray}
where $\xi$ is the radial coordinate for a spherical or cylindrical cell
of radius $R$, or the distance from the central plane of a slab of
half-size $R$ ($\xi=r$, $r_\perp$, or $|z|$ for the three respective
cases). The function $f_q(\xi)$ describes the shape of the density profile,
which can depend on adjustable parameters, and $\hat{f}_q(\xi)$ is the
normalized density profile, so that the parameter ${n}_q^\mathrm{out}$
has the meaning of a nucleon density outside the ``nuclei'' (the
nucleonic clusters), 
and 
more generally at the cell boundary. In the bulk
of the inner crust, free protons are absent and the nucleons are
clustered near the center of the WS cell; in this case
${n}_{p}^\mathrm{out}=0$. However, free protons may appear near the 
transition to the core of the star \cite{Pearson_18}. We assume
$f_q(0)>f_q(R)\geq0$, 
while ${n}_{\Lambda q}$ can be positive for the normal
phases or negative for the inverse phases.  Accordingly, ${n}_{\Lambda
q} = {n}_q^\mathrm{cen}-{n}_q^\mathrm{out}$ is the central number
density excess,  ${n}_q^\mathrm{cen}$ being the density at the center of
the cell. Hereafter we will mark different parametrizations of
$f_q(\xi)$ by abbreviations in superscripts.

Oyamatsu \cite{Oyamatsu93} 
studied nuclear shapes in the neutron star mantle using the 
Bethe theory \cite{Bethe68} and assuming the shapes of
neutron and proton number density distributions in the form
\begin{equation}
  f_q^\mathrm{O}(\xi) = \left\{
    \begin{array}{lll}
        \left[ 1 - \left( { \xi / r_q }
        \right)^{t_q} \right]^3
       \mbox{~for~}  \xi < r_q, \\[1ex]
       0    
       \mbox{~~for~} \xi \geq r_q,
     \end{array}
        \right.
\label{Oya}
\end{equation}
An analogous parametrization, but with
$t_n=t_p$, $r_n=r_p$, and with the power index 2 instead of 3, was
previously used by Arponen \cite{Arponen72}.
Parameters $t_q$ control the sharpness of the local
density profiles, while $r_q$ 
determine the neutron and proton radii of a nucleus ($0 < r_q < R$).
In this case $\hat{f}_q^\mathrm{O}(\xi)=f_q^\mathrm{O}(\xi)$.
With increasing density
$\bar{n}$, 
the 
profiles become smoother, approaching the limit of
uniform matter; therefore, the parameters $t_q$ decrease.

The parametrization (\ref{Oya}) was widely used in TF
and second-order ETF calculations
\cite{GogeleinMuther07,LimHolt17,OyamatsuIida07,OyamatsuIida10}. 
However, 
real local density distributions of neutrons and protons in a neutron
star crust are not cut off at a certain distance from the center of a WS
cell. Therefore,  $r_{n}$ and $r_{p}$ can be treated only as convenient
fit parameters. Near the bottom of the crust, the local density
distribution is rather smooth, and the boundary between the free and
bound neutrons becomes rather uncertain. 
More importantly, 
this parametrization
leads to divergences for the fourth-order ETF energy.

In a number of applications of the ETF methods to finite nuclei (with
${n}_q^\mathrm{out}=0$; e.g., \cite{Dutta_86}) and neutron-star mantles
(e.g., \cite{OnsiPP97,ShelleyPastore21}), the simple Fermi-function form
(also known as the Woods-Saxon shape
\cite{BarrancoBuchler81,GogeleinMuther07}) was adopted, which can be
written as
\begin{equation}
  f_q^\mathrm{F}(\xi) =
 \frac{1}{1 +\exp \Big(\frac{\xi-r_q}{a_q}\Big) } ,
\label{Fermi}
\end{equation}
where $r_q$ is the half-height nuclear radius and $a_q$ accounts for the
diffuseness of the  nuclear surface. This parametrization was
criticized on the grounds that it cannot capture the
asymmetry of nucleon density profiles at the surface of a
nucleus \cite{Suraud85}. To overcome the symmetry constraint, a
modification of
Eq.~(\ref{Fermi}) was used in a number of
ETF calculations \cite{Vinas_84,Kolehmainen_85,Pi_86,MartinUrban15},
\begin{equation}
f_q^\mathrm{MF}(\xi) =
 \frac{1}{\left[ 1 +\exp
 \Big(\frac{\xi-r_q}{a_q}\Big)\right]^{\nu_q} } ,
\label{MF}
\end{equation}
where the power index $\nu_q$ is an additional fit parameter. Still more
general modification, which allows for an additional enhancement or
depression of the nucleon density at the center of the WS cell, was
considered in Refs.~\cite{ChuJB77,BrackGH85}, but the minimized energy
was found to be insensitive to the latter degree of freedom. Meanwhile,
an increase of the power index $\nu_q$ from 1 to 3 decreases the
calculated energy in
a heavy nucleus by $\sim8$~MeV \cite{ChuJB77}, although it does not
substantially change the results for the inner crust and mantle of a
neutron star~\cite{MartinUrban15}.

It is easy to see that neither $f_q^\mathrm{F}$ nor $f_q^\mathrm{MF}$
can provide trial functions $n_q(\bm{r})$ 
satisfying 
condition
(\ref{gradS}). Below we will consider other modifications of the
Fermi-like parametrization (\ref{Fermi}), which can be written in the
generic form
\begin{equation}
f_q(\xi) = \frac{1}{1 + 
h(\xi;r_q,a_q) \exp \Big(\frac{\xi-r_q}{a_q}\Big) } .
\label{fh}
\end{equation}
The first example of such parametrization with
\begin{equation}
h^\mathrm{DF}(\xi;r_q,a_q) = \exp \left[\Big(\frac{r_q - R}
{\xi - R}\Big)^2 - 1\right]
\label{DF}
\end{equation}
was introduced by Onsi et al.~\cite{Onsi_08}. The resulting ``damped
Fermi parametrization'' $f^\mathrm{DF}_q(\xi)$, which has \emph{all}
derivatives vanishing at $\xi=R$, was consistently used for trial
density profiles in the calculations of the properties of neutron-star
inner crust and mantle by the Brussels-Montreal group (e.g.,
Ref.\,\cite{ShchechilinCP23}, and references therein).

Density profiles ${n}_q(\bm{r})$ parametrized with the shape forms
$f_q^\mathrm{F}(\xi)$, $f_q^\mathrm{MF}(\xi)$, and
$f_q^\mathrm{DF}(\xi)$ do not have well defined gradients at the centers
of the WS cells. This fact is often neglected by considering only the
interval 
$\xi\in(0,R]$ 
without inclusion the origin. The gradient
corrections can be defined in the entire WS cell, if $n_q(\bm{r})$ is
treated as a generalized distribution or is meant to be locally smoothed
in a negligibly small neighborhood of the center \cite{PearsonCP20}.
Nevertheless such parametrizations can cause potential problems,
discussed in the next sections. Besides, in the case of the ``lasagna''
phase the non-zero derivative of ${n}_q(z)$ at $z\to0$ hampers the
aforementioned replacement of $\tau_q^{(4)}$  by $\tau_{q,v}^{(4)}$ in
Eq.~(\ref{Tq}), because $\mathcal{T}_{q,s}^{(4)}$ does not vanish at the
surface $z=0$.

In the latter (``lasagna'') case, there was also a symmetry argument
disfavoring parametrization (\ref{DF}). Unlike the distinction between
the normal quasi-spherical nuclei and ``Swiss cheese'' or between
``spaghetti'' and ``anti-spaghetti'' phases, there is no physical
distinction between ``lasagna'' and ``anti-lasagna'': a configuration
with a maximum at the center of the WS cell can be transformed into a
configuration with a minimum at the center by simple translation of the
coordinate system. In practice, distinct lasagna and anti-lasagna
parametrizations can give very close energy minima producing a spurious
instability of a numerical minimization procedure. To avoid it, both
lasagna and anti-lasagna should be described equally well by the
chosen parametrization.  This  means that for any profile
${n}_q(\xi)$, determined by a parameter set $\bm{\chi}$, there should
exist a set of parameters $\bm{\chi'}$ such that the inverted profile
${n}_q(R-\xi)$ obtained from a space inversion followed by a 
translation is also allowed \cite{Shchechilin_24}\footnote{
Alternatively, one can avoid the lasagna-anti-lasagna distinction by
imposing the condition ${n}_{\Lambda q} > 0$.}:
\begin{equation}
\forall\bm{\chi}~\exists\bm{\chi'}: \quad {n}_q(\xi,\bm{\chi})={n}_q(R-\xi,\bm{\chi'}) \, .
\label{lasagna_req}
\end{equation}
If ${n}_q(\xi,\bm{\chi})$ satisfies condition (\ref{gradS})
for any $\bm{\chi}$, then condition (\ref{lasagna_req}) ensures that
${d}{n}_q/{d}\xi \to0$ at $\xi\to0$, so that the gradient of the trial
density distribution does exist and equals zero at the WS cell center.

Parametrizations of the form~(\ref{fh}) are symmetric with respect to 
lasagna and anti-lasagna configurations provided 
\begin{equation}
h(\xi;r_q,a_q) = \frac{1}{h(R - \xi; R-r_q, a_q)} .
\label{h_inv}
\end{equation}
The constraint~(\ref{lasagna_req}) is then satisfied with 
${n}_{\Lambda q}^\prime=-{n}_{\Lambda q}$, ${{n}_q^\mathrm{out}}^\prime
={n}_q^\mathrm{out}+{n}_{\Lambda q}$,  $a_q^\prime=a_q$, and
$r_q^\prime=R-r_q$. 

The particular form $h^\mathrm{DF}(\xi;r_q,a_q)$ in
Eq.~(\ref{DF})  does not satisfy the condition (\ref{h_inv}). A
straightforward generalization to fulfill this condition reads 
\begin{equation}
h^\mathrm{2DF}(\xi;r_q,a_q) = \exp \left[\Big(\frac{r_q - R}
{\xi - R}\Big)^2 - \Big(\frac{r_q}
{\xi}\Big)^2\right].
\label{2DF}
\end{equation}
In particular, it ensures that all derivatives of ${n}_q(\bm{r})$ vanish not only at
the boundary, but also at the center of the WS cell. However, a
comparison of the ETF calculations with different density profile
parametrizations \cite{Shchechilin_24} indicates that even the one-sided
``strong damping'' defined by Eq.~(\ref{DF}), let alone the two-sided
damping in Eq.~(\ref{2DF}), is
too restrictive, because a noticeably lower minimum of $E$ can be obtained with a 
weaker
(``soft'') damping, defined by
\begin{equation}
h^\mathrm{WDF}(\xi;r_q,a_q) = 
\left(\frac{r_q-R} {r_q}\right)^2 \left(\frac{\xi}{\xi-R}\right)^2 .
\end{equation}
The corresponding density profile not only satisfies condition
(\ref{gradS}), but also has zero gradient at the center of the WS cell.  
Meanwhile, higher-order density derivatives do not vanish at the WS cell
center or boundary. 
In the next section we will show that the parametrizations, which do not
have the zero gradient at $\xi=0$, lead to a divergence of the ETF energy
functional $E$ in the cylindrical and plane-parallel symmetries.


\section{Divergences of non-smooth parametrizations}
\label{sect:divergence}

\subsection{Symmetric parametrizations in pasta phases}

Let us assume that $f_q(\xi)$ is finite and differentiable
at $\xi > 0$. Then we can write
\begin{equation}
   {n}_q(\bm{r}) = {n}_{q0} + {n}_{q0}' \,\xi + 
   O(\xi^2),
\label{Taylor}
\end{equation}
where $n_{q0} = {n}_q^\mathrm{out} + {n}_{\Lambda q}$ and 
\[
{n}_{q0}' =
{n}_{\Lambda q} \frac{{d}\hat{f}_q(\xi)}{{d}\xi}\bigg|_{\xi\to+0}.
\]
If ${n}_{q0}'\neq0$, then the distribution ${n}_q$
has a kink at the center of the cell, and
$\nabla{n}_q$ is not well defined at this point.

Let us evaluate a contribution to the integral (\ref{Tq}) from the first
term on the right-hand side of Eq.~(\ref{L4}),
\begin{equation}
   \tau_q^{(4a)} = 
 \frac{\hbar^2}{2M_q}\,\frac{{n}_q^{1/3}}{(3\pi^2)^{2/3}} \,
     \frac{1}{270}\left(\frac{\nabla^2{n}_q}{{n}_q}\right)^2
     \propto
     {n}_q^{1/3}\left(\frac{\nabla^2{n}_q}{{n}_q}\right)^2   ,
\label{prop}
\end{equation}
in a small neighborhood $\xi<\varepsilon$ 
of the center of the cell.
In the case of cylindrical or plane-parallel symmetry, 
we will consider WS cells of finite volume $V$, respectively 
cylinders of unit length and slabs of unit area. The volume of 
the $\varepsilon$-neighborhood of their center will be denoted 
by $V_\varepsilon$.
For the plane-parallel, cylindrical, or
spherical symmetry, $\nabla^2{n}_q =
\xi^{1-D}\,({\partial}/{\partial \xi})
     (\xi^{D-1} {\partial {n}_q}/{\partial \xi} )$,
where $D = 1,$ 2, or 3, respectively.

In the case of the spherical symmetry, Eq.~(\ref{Taylor}) gives
\begin{eqnarray}&&\hspace*{-2em}
   \int_{V_\varepsilon} {n}_q^{1/3}
   \left(\frac{\nabla^2{n}_q}{{n}_q}\right)^2\,{d} V
= 4\pi
   \int_0^\varepsilon {n}_q^{-5/3}
  \left(
     \nabla^2{n}_q \right)^2
     r^2\,{d} r 
\nonumber\\&&
= 16\pi \,{n}_{q0}^{-5/3}\,({n}_{q0}')^2 \, \varepsilon
   + O(\varepsilon^2).
\end{eqnarray}
Here, the right-hand side tends to zero at $\varepsilon\to0$.
Therefore the kink at $r=0$
can be safely isolated
by removing a sphere of sufficiently small radius $\varepsilon$
around the center
without an appreciable effect on the integral~(\ref{Tq}).

In the case  of cylindrical symmetry,
\begin{eqnarray}&&\hspace*{-2em}
   \int_{V_\varepsilon}\!\! {n}_q^{1/3}
   \left(\frac{\nabla^2{n}_q}{{n}_q}\right)^2\,{d} V
= 2\pi
   \int_0^\varepsilon {n}_q^{-5/3}
  \left(
     \nabla^2{n}_q \right)^2
     r_\perp\,{d} r_\perp
\nonumber\\&&
= 2\pi {n}_{q0}^{-5/3}\,({n}_{q0}')^2
   \int_0^\varepsilon
      \frac{{d} r_\perp}{r_\perp} 
   + O(\varepsilon).
\end{eqnarray}
The last integral diverges.
Therefore, the contribution of this gradient correction to
$E_\mathrm{nuc}$
is infinite, unless ${n}_{q0}'=0$.
This result shows that the energy of a cylindrical WS cell (per unit length of the
cylinder),
calculated according to the ETF theory, can be made finite
only by such a density distribution ${n}_q(r_\perp)$
that
\begin{equation}
   \lim_{r_\perp\to0}\,\frac{{d} {n}_q}{{d} r_\perp} = 0.
\label{cyl0cond}
\end{equation}

For the plane-parallel WS cells, we have $\xi=|z|$.
Taking into account that
$
   {{d} \,|z|}/{{d} z} = 2\theta(z)-1,
$
where $\theta(z)$ is the Heaviside step function,
and
$
   {{d} \theta(z)}/{{d} z} = \delta(z),
$
where $\delta(z)$ is the Dirac delta function,
we obtain
$
   \nabla^2{n}_q = 2{n}_{q0}'\,\delta(z) + O(1).
$
Now
\begin{equation}
   \int_{V_\varepsilon} {n}_q^{1/3}
   \left(\frac{\nabla^2{n}_q}{{n}_q}\right)^2 {d} V
  =\frac{(2{n}_{q0}')^2}{{n}_{q0}^{5/3}}
   \int_{-\varepsilon}^\varepsilon 
      [\delta(z)]^2\,{d} z
      + O(1).
\label{planediv}
\end{equation}
Since the squared delta function is not integrable, the integral is
finite only if ${n}_{q0}'=0$, that is
\begin{equation}
   \lim_{z\to0}\frac{{d} {n}_q}{{d} z} = 0.
\label{pla0cond}
\end{equation}


\subsection{Continuity across a surface}

Condition (\ref{pla0cond}) can be generalized. Let us consider
a continuous distribution ${n}_q(\bm{r})$ which 
has a kink (i.e., a discontinuous first derivative) at the plane $z=0$:
${n}_q(\bm{r}) = {n}_{q0}(x,y) + {n}_+'(x,y) \,z + O(z^2)$ at $z>0$ and
${n}_q(\bm{r}) = {n}_{q0}(x,y) + {n}_-'(x,y) \,z + O(z^2)$ at $z<0$. It can be written as
\begin{eqnarray}
   {n}_q(\bm{r}) &=& {n}_{q0}(x,y) + \frac{{n}_+'(x,y) + {n}_-'(x,y)}{2}\,z
\nonumber\\&&
          + \frac{{n}_+'(x,y) - {n}_-'(x,y)}{2}\,|z| + O(z^2).
\end{eqnarray} 
By analogy with Eq.~(\ref{planediv}),
a contribution  of the term containing $|z|$ to the integral of $\tau_q^{(4a)}$
over an $\varepsilon$-neighborhood $V_\varepsilon = \Delta S \otimes
[-\varepsilon,\varepsilon]$
 of any finite element $\Delta S$ of the plane $z=0$ (for example,
parallelepiped
$V_\varepsilon = \{x,y,z: |x|< \Delta x, |y| < \Delta y, |z| <
\varepsilon \}$) is proportional to
\begin{equation}
   \int_{V_\varepsilon} {n}_q^{1/3}
   \left(\frac{\nabla^2{n}_q}{{n}_q}\right)^2 {d} V
  = A_{\Delta S} 
   \int_{-\varepsilon}^\varepsilon 
      [\delta(z)]^2\,{d} z
      + O(1),
\label{planegen}
\end{equation}
where 
\begin{equation}
A_{\Delta S} = \int_{\Delta S} \frac{[{n}_+'(x,y) -
{n}_-'(x,y)]^2}{{n}_{q0}^{5/3}(x,y)} \,{d}S.
\end{equation}
If $A_{\Delta S} \neq 0$, the integral (\ref{planegen}) diverges.
Since the surface element $\Delta S$ is arbitrary, the divergence can be excluded only if 
${n}_+'(x,y) \equiv {n}_-'(x,y)$, except possibly a zero-measure submanifold of
points $(x,y)$. This means the absence of a kink across the
plane.

If ${n}_q(\bm{r})$ has a kink at a smooth surface, which is not plane,
one can approximate a sufficiently small area around an arbitrary point
on this surface by the tangent plane and apply the above considerations
to prove the divergence.    Thus we conclude that ${n}_q(\bm{r})$ should
not have a kink on a smooth surface. As a particular case, it should not
have a kink on the WS cell boundary, if instead of spheres and cylinders
one considers the true polyhedral WS cells filling the space and
minimizes $E$ in two or more adjacent cells together. Assuming that
${n}_q(\bm{r})$ is symmetric and periodic, this leads to 
Eq.~(\ref{gradS}).

Although we have considered only one term in Eq.~(\ref{L4}),  obtained
with the use of the Ostrogradsky-Gauss theorem, the drawn conclusions
are general and do not assume fulfillment of condition (\ref{gradS}) in
advance. Indeed, we can isolate a kink location by a surface, across which
$n_q(\bm{r})$ is smooth, and apply this theorem only to the contribution
into the integral (\ref{Tq4}) from the domain surrounded by this
surface. Then the volume integral of $\tau_{q,v}^{(4)}$ over the
considered domain contains the above-discussed divergence, whereas the
corresponding surface integral is finite and cannot cancel it. Besides,
the third term in the original expression (\ref{T4}), being proportional
to $\tau_q^{(4a)}$ (\ref{prop}), contains the same 
divergence.


\section{Accuracy limit for non-smooth parametrizations}
\label{sect:acc_lim}

One can try to circumvent the smoothness conditions, such as
Eq.~(\ref{cyl0cond}) or Eq.~(\ref{pla0cond}), by introducing a local
modification of  ${n}_q(\bm{r})$ in a small neighborhood $V_\varepsilon$
around the origin
so that
the modified  trial density distribution $\tilde{n}_q(\bm{r})$ coincides
with ${n}_q(\bm{r})$ outside $V_\varepsilon$ and has a continuous gradient
everywhere. Since $\varepsilon$ is small and $\tilde{n}_q$ is symmetric
and differentiable, we have $\tilde{n}_q|_{\xi\leq\varepsilon} =
{n}_{q0} + O(\varepsilon)$ and
$\nabla\tilde{n}_q|_{\xi=\varepsilon}\cdot\mathbf{n}_{\varepsilon} =
{n}_{q0}' + O(\varepsilon)$, where $\mathbf{n}_{\varepsilon}$ is the
outer normal to the surface of $V_\varepsilon$, which we denote
${S_\varepsilon}$. Then  the average of $\nabla^2\tilde{n}_q$ in
$V_\varepsilon$ equals
\begin{eqnarray}
\left\langle \nabla^2\tilde{n}_q \right\rangle
&\equiv&
  \frac{1}{||V_\varepsilon||} 
   \int_{V_\varepsilon}\!\! \nabla^2\tilde{n}_q\,{d} V
 =  
   \oint_{S_\varepsilon}\!\!
   \nabla\tilde{n}_q|_{\xi=\varepsilon}\cdot\mathbf{n}_{\varepsilon}
   \frac{{d}S}{||V_\varepsilon||}
\nonumber\\&=&
   \left[{n}_{q0}'  + O(\varepsilon)\right]\frac{||S_\varepsilon||}{||V_\varepsilon||}
   =
   \frac{{n}_{q0}' D}{\varepsilon} + O(1),
\label{nabla2n}
\end{eqnarray}
where $||V_\varepsilon|| = ||S_\varepsilon||\,\varepsilon/D =
\pi^{D/2}\,\varepsilon^D/\Gamma(D/2+1)$ 
and the angle brackets denote
the averaging.
Assuming that ${n}_{q0} \neq 0$ and using the inequality 
$\langle x^2 \rangle \geq \langle x \rangle^2$, we obtain
\begin{eqnarray}&&\!\!\!\!
\int_{V_\varepsilon} \frac{ \left(\nabla^2\tilde{n}_q\right)^2
     }{\tilde{n}_q^{5/3}}
     \,{d} V
  \approx
     \frac{||V_\varepsilon||}{{n}_{q0}^{5/3}}
     \left\langle \left(\nabla^2\tilde{n}_q\right)^2 \right\rangle
\nonumber\\&&\qquad
     \geq
     \frac{||V_\varepsilon||}{{n}_{q0}^{5/3}}
     \left\langle \nabla^2\tilde{n}_q \right\rangle^2
  \approx
     \frac{D^2||V_\varepsilon||}{\varepsilon^2 {n}_{q0}^{5/3}}\,|{n}_{q0}'|^2,
\end{eqnarray}
where the approximate equality implies an accuracy up to the factor of
$1+O(\varepsilon)$.
Thus, in the limit of small $\varepsilon$,
the leading contribution of the integral of $\tau_q^{(4)}$ over $V_\varepsilon$
to the kinetic energy correction $\mathcal{T}_q^{(4)}$ can be bounded
from below as 
\begin{equation}
\mathcal{T}_{q,\varepsilon}^{(4a)} \equiv \int_{V\varepsilon}
\tau_q^{(4a)} \,{d}V \gtrsim
 \frac{\hbar^2}{540M_q}\,\frac{D^2||V_\varepsilon||}{(3\pi^2)^{2/3}} \,
 \frac{|{n}_{q0}'|^2}{\varepsilon^2 {n}_{q0}^{5/3}}\,.
\label{T4eps}
\end{equation}
When considering a WS cell fragment in the lasagna phase ($D=1$),
Eq.~(\ref{T4eps}) leads to the following inequality for
the contribution of the kink to the kinetic energy per one nucleon:
\begin{equation}
\frac{\mathcal{T}_{q,\varepsilon}^{(4a)}}{N} \gtrsim 
 \frac{\hbar^2}{540M_q}\,\frac{1}{(3\pi^2)^{2/3}} \,
\frac{ {n}_{q0}^{1/3}}{\bar{n}R}
\left(\frac{{n}_{q0}'}{{n}_{q0}}\right)^{\!\!2}
\varepsilon^{-1}.
\label{plane_est}
\end{equation}
The right-hand side of Eq.~(\ref{plane_est})
 tends to infinity  at $\varepsilon\to0$, unless ${n}_{q0}'=0$.
Thus, an accuracy of a discrete-mesh calculation of the
fourth-order ETF energy has a fundamental limit for non-smooth trial
functions, because the mesh cannot be arbitrarily refined: the smaller
the interval $(-\varepsilon,\varepsilon)$, over which the smoothing is
performed, the larger its contribution.

For example, the pasta phases were studied in Ref.~\cite{PearsonCP20}
using the BSk24  nuclear energy-density functional with the damped Fermi
(DF) parametrization, defined by Eqs.~(\ref{fh}) and (\ref{DF}). The
lasagna phase was found to exist in the density range
$\bar{n}\sim(0.07-0.08)$~fm$^{-3}$. In this range, the optimal
variational
parameters were, respectively, $R\approx (13.7-13.0)$~fm, $r_p\approx
(4.7-6.6)$~fm, $r_n-r_p\sim 0.5$~fm, $a_{n,p}\approx(1.0-1.3)$~fm,
${n}_{\Lambda n}\approx(0.028-0.019)$~fm$^{-3}$,
${n}_n^\mathrm{out}\approx(0.058-0.065)$~fm$^{-3}$, and ${n}_{\Lambda
p}\approx(0.007-0.005)$~fm$^{-3}$. Then Eqs.~(\ref{nLambda}),
(\ref{fnorm}), (\ref{fh}), (\ref{DF}), and (\ref{plane_est}) give
$(\mathcal{T}_{n,\varepsilon}^{(4a)}+\mathcal{T}_{p,\varepsilon}^{(4a)})/{N}
\gtrsim (0.034-0.007)\mbox{\,eV~fm}/\varepsilon$. In the considered
density range, energies of different pasta phases differ typically by a
few hundreds of eV per nucleon. Therefore the accuracy limitation would be essential
when performing numerical integration of the ETF equations with steps
$\varepsilon \lesssim10^{-3}$~fm. In fact, a much coarser mesh was used
in Ref.~\cite{PearsonCP20}, which provided the practical
imperceptibility of the results of that study to the accuracy limit
(\ref{plane_est}).


\section{Conclusions}
\label{sect:concl}

We considered  the smoothness conditions on trial density distribution
${n}_q(\bm{r})$ for the applicability of the fourth-order ETF theory in
the approximations of spherical, cylindrical, and plane-parallel WS
cells. We have shown that a gradient discontinuity (a kink) of a trial 
density distribution at any smooth surface (which can be a WS cell
surface) or at the center of a cylindrical or plane-parallel WS
cell makes the fourth-order ETF gradient correction
divergent. On the other hand, a kink of ${n}_q(\bm{r})$
at the center of a spherical WS cell does not result in a divergence. In
the latter case, the second and fourth order gradient corrections can be
treated as distributions and remain integrable.

In previous discrete-mesh calculations, a kink of ${n}_q(\bm{r})$
was sometimes understood (e.g., Ref.~\cite{PearsonCP20}) as
an approximation to a sufficiently smooth function $\tilde{n}_q(\bm{r})$, which was not explicitly defined
but was meant to coincide with the employed trial function
${n}_q(\bm{r})$ at the mesh nodes. In Sec.~\ref{sect:acc_lim} we
demonstrated a fundamental limitation of the accuracy inherent to this
approach. 

This accuracy limitation, however, does not invalidate the previously
reported results \cite{PearsonCP20,ShchechilinCP23}, because it lies
well below their actual accuracy. According to these calculations, in
order to correctly determine the structural phase transitions between
the nuclear pasta phases, it is sufficient to calculate energy per
nucleon with an accuracy of $\lesssim0.1$~keV. Meanwhile, the minimal
contribution of the smoothed-out central kink in the DF
parametrization (\ref{DF}),  evaluated according to Eq.~(\ref{plane_est}), does not
exceed 0.1 keV per nucleon in the lasagna phase, if
$\varepsilon\gtrsim10^{-3}$~fm. As mentioned in Ref.~\cite{PearsonCP20},
the integrals calculated  with a mesh size of 0.1~fm correspond to the
kink having been smoothed out locally over the region
$\xi\lesssim0.01$~fm, which is sufficient for the required accuracy. 

We should emphasize that the above numerical example pertains to the concrete
implementation of the ETF theory, based on the BSk24 energy density
functional \cite{PearsonCP20}. Besides, we have used the approximation
$M_q^* = M_q$, which means that our numerical estimates are only valid by
order of magnitude, but not exact. The limit on the step in $\xi$ can be different with
another nuclear interaction model. The result can also depend on the
interval of the scanned parameters in the minimization procedure. For
instance, because of the divergences, a numerical minimization can
sometimes result in unrealistic values of parameters with much bigger
error on the energy in comparison with optimal parameters. 

Anyway, the implicit smoothing-out of the divergences entails a
risk of spoiling the results through the dependence on the rather
arbitrary smoothing function. For example, a numerical integration of
the kinetic energy density with an automated choice of an integration
step might occasionally ``feel'' the contribution from the kink, which
would result in unlimited refinement of the mesh to ever smaller step
sizes $\varepsilon$ near the kink points, thus leading to the divergence
evaluated in Sec.~\ref{sect:acc_lim}. Thus we conclude that using
trial functions with continuous first derivatives everywhere, including
the center of a WS cell, should be recommended for solving the
fourth order ETF equations by the restricted variational method.\\

A.P.\ thanks D.G.~Yakovlev for useful remarks on a preliminary text of
the paper. The work of A.P.\ and A.C.\ was supported by the Russian
Science Foundation grant No.\,22-12-00048.
The work of N.S. was financially supported by the FWO (Belgium) and the 
Fonds de la Recherche Scientifique (Belgium) under the Excellence of Science 
(EOS) programme (project No.\,40007501). The work of N.C. received funding from 
the Fonds de la Recherche Scientifique (Belgium) under Grant No.~IISN~4.4502.19.


\newcommand{\artref}[4]{\textit{#1} \textbf{#2} #3 (#4)}
\newcommand{\aap}{Astron.\ Astrophys.}
\newcommand{\aaps}{Astron.\ Astrophys.\ Suppl.\ Ser.}
\newcommand{\apj}{Astrophys.\ J.}
\newcommand{\mnras}{Mon.\ Not.\ R.\ Astron.\ Soc.}
\newcommand{\physrep}{Phys.\ Rep.}
\newcommand{\plb}{Phys.\ Lett. B}
\newcommand{\prl}{Phys.\ Rev.\ Lett.}
\newcommand{\prc}{Phys.\ Rev.~C}
\newcommand{\nphysa}{Nucl.\ Phys.\ A}

\renewcommand{\refname}{References}
\makeatletter
\renewcommand\@biblabel[1]{#1.}
\makeatother

\end{document}